\documentclass[12pt,a4paper]{article}

\usepackage{amsfonts,amssymb}
\usepackage{amsmath}
\usepackage{cleveref}

\newcommand{\EQ}{\begin{equation}}
\newcommand{\EN}{\end{equation}}

\newtheorem{theorem}{Theorem}[section]
\newtheorem{corollary}{Corollary}
\newtheorem{lemma}[theorem]{Lemma}
\newtheorem{proposition}{Proposition}

\newtheorem{definition}[theorem]{Definition}
\newtheorem{remark}{Remark}

\newcommand{\zero}{{\mathbf{0}}}

\newcommand{\bc}{{\bf c}}

\newcommand{\by}{{\bf y}}
\newcommand{\bx}{{\bf x}}

\newcommand{\bv}{{\bf v}}

\newcommand{\F}{\mathbb{F}}

\newcommand{\al}{\alpha}
\newcommand{\be}{\beta}
\newcommand{\ga}{\gamma}

\newcommand{\IA}{\operatorname{IA}}

\newcommand{\Aut}{\operatorname{Aut}}
\newcommand{\MAut}{\operatorname{MAut}}
\newcommand{\GL}{\operatorname{GL}}
\newcommand{\pr}{\indent{\bf Proof. \ }}
\newcommand{\qed}{\hspace*{5 mm}$\Box$}

\title{On new infinite families of completely regular and completely
transitive codes}

\author{J. Borges$^{*}$ \and J. Rif\`{a}\thanks{Department of Information and Communications Engineering. Universitat Aut\`{o}noma de Barcelona.} \and V. A. Zinoviev\thanks{A.A. Kharkevich Institute for Problems of Information Transmission. Russian Academy of Sciences.}
}

\begin{document}
\maketitle

\bigskip

\begin{abstract}
In two previous papers we constructed new families of completely
regular codes by concatenation methods. Here we determine cases in which the new codes are completely transitive.
For these cases we also find
the automorphism groups of such codes.
For the remaining cases, we show that the codes are not completely transitive assuming un upper bound on the order of the monomial automorphism groups, according to computational results.
\end{abstract}

\section{Introduction}
Let $\F_q$ be the finite field of order $q$, where $q$ is a prime power. For a $q$-ary {\em code} $C\subset \F_q^n$ of {\em length} $n$, denote by $d$ its {\em minimum (Hamming) distance} between any pair of distinct codewords. The {\em packing radius} of $C$ is
 $e=\lfloor (d-1)/2 \rfloor$. Given any vector $\bv \in \F_q^n$, its
{\em distance to the code $C$} is $d(\bv,C)=\min_{\bx \in C}\{
d(\bv, \bx)\}$ and the {\em covering radius} of the code $C$ is
$\rho=\max_{\bv \in \F_2^n} \{d(\bv, C)\}$. Note that $e\leq \rho$.
A {\em linear} $[n,k,d;\rho]_q$-{\em code} is a $k$-dimensional subspace of $\F_q^n$, with minimum distance $d$ and covering radius $\rho$. We denote by $~D=C+\bx~$ ($\bx\in\F_q^n$) a {\em coset} of  $C$, where $+$ means the component-wise addition in $\F_q$.

For a given code $C$ of length $n$ and covering radius $\rho$,
define
\[
C(i)~=~\{\bx \in \F_q^n:\;d(\bx,C)=i\},\;\;i=0,1,\ldots,\rho.
\]
The sets $C(0)=C,C(1),\ldots,C(\rho)$ are called the {\em subconstituents} of $C$.

Say that two vectors $\bx$ and $\by$ are {\em neighbors} if $d(\bx,\by)=1$. Denote by $\zero$ the all-zero vector or the all-zero matrix, depending on the context.

\begin{definition}[\cite{Neum}]\label{de:1.1} A code $C$ of length
$n$ and covering radius $\rho$ is {\em completely regular}, if for all $l\geq 0$ every vector $x \in C(l)$ has the same
number $c_l$ of neighbors in $C(l-1)$ and the same number $b_l$ of
neighbors in $C(l+1)$. Define $a_l = (q-1){\cdot}n-b_l-c_l$ and
set $c_0=b_\rho=0$. The parameters $a_l$, $b_l$ and
$c_l$ ($0\leq l\leq \rho$) are called {\em intersection numbers}
and the sequence $\{b_0, \ldots, b_{\rho-1}; c_1,\ldots, c_{\rho}\}$
is called the {\em intersection array} (shortly $\IA$) of $C$.
\end{definition}

Let $M$ be a monomial matrix, i.e. a matrix with exactly one
nonzero entry in each row and column. If $q$ is prime, then the
automorphism group of $C$, $\Aut(C)$, consists of all monomial
($n\times n$)-matrices $M$ over $\F_q$ such that $\bc M \in C$ for
all $\bc \in C$. If $q$ is a power of a prime number, then
$\Aut(C)$ also contains any field automorphism of $\F_q$ which
preserves $C$. The group $\Aut(C)$ acts on the set of cosets of
$C$ in the following way: for all $\pi\in \Aut(C)$ and for every
vector $\bv \in \F_q^n$ we have $\pi(\bv + C) = \pi(\bv) + C$.
Fix the following notation for groups, which we need: let $C_t$ denote the cyclic group of order $t$, let $S_t$ denote the symmetric group of order $t!$, and let $\GL(m,q)$ be the $q$-ary general linear group (formed by all nonsingular $q$-ary $(n \times n)$-matrices).

%

In this paper, we consider only monomial automorphisms. Thus, when $q$ is not a prime, but a prime power, we omit the field automorphisms that fix the code. We denote by $\MAut(C)$ the monomial automorphism group of a code $C$.

\begin{definition}\label{de:1.3}
Let $C$ be a  $q$-ary linear code with covering radius $\rho$.
Then $C$ is {\em completely transitive} if $\MAut(C)$ has $\rho +1$ orbits
when acts on the cosets of $C$.
\end{definition}

Note that this definition generalizes to the $q$-ary case the definition
given in \cite{Sole}. However, it is a particular case of the definition
of coset-completely transitive code given in \cite{Giu}, where not only
monomial automorphisms are considered, but also the field automorphisms
are taken into account.

Since two cosets in the same orbit have the same weight
distribution, it is clear that any completely transitive code is
completely regular.

It is well known, e.g. see \cite{MacW}, that the monomial automorphism
group of a Hamming code $\mathcal{H}_m$ of length $(q^m-1)/(q-1)$ is
isomorphic to the general linear group $\GL(m, q)$, which acts transitively
on the set of vectors of weight one. As a consequence, all cosets of minimum
weight one are in the same orbit and thus Hamming codes are completely
transitive. In the binary case, the action of $\GL(m, 2)$ on the set of
coordinate positions is even doubly transitive (but only in the binary
case). Later, we will use this fact.

Completely regular and completely transitive codes are classical
subjects in algebraic coding theory, which are closely connected
with graph theory, combinatorial designs and algebraic
combinatorics. Existence, construction and enumeration of all such
codes are open hard problems (see \cite{BCN,Dam,Koo,Neum,BRZ5} and
references there).

It is well known that new completely regular codes can be obtained
by the direct sum of perfect codes or, more general, by the direct
sum of completely regular codes with covering radius $1$
\cite{BZZ,Sole}. In the previous papers \cite{BRZ3, BRZ4}, we
extend these constructions, giving several explicit constructions
of new completely regular codes, based on concatenation methods.
Here, we find the monomial automorphism groups for that families
in some cases.  This gives mutually nonequivalent binary linear
completely regular codes with the same intersection arrays and
isomorphic monomial automorphism groups. We show cases in which
the constructed families of codes are also completely transitive.

In \Cref{sec:2} we give some preliminary results, mostly coming
from the previous papers \cite{BRZ2, BRZ3, BRZ4} (where we
introduced two infinite families of CR codes $A^{(r)}$ and
$B^{(r)}$), that help to us place ourselves in the problem we want
to address. In \Cref{sec:3} we study the completely transitive
character of codes $B^{(r)}$ and calculate $\MAut(B^{(r)})$ for
some cases. Finally, we give a result that gives us the character
of completely transitive for these $B^{(r)}$ codes with an extra
assumption about the order of $\MAut(B^{(r)})$. This assumption is
given by computational considerations on the order of such
monomial automorphism groups.

\section{Preliminary results}\label{sec:2}

In this section we recall the results of \cite{BRZ2, BRZ3, BRZ4}. Combining such results, in the next section, we specify cases in which the constructed infinite families of completely regular codes are also completely transitive.

For any vector $x=(x_1,\ldots,x_n)\in \F_q^n$, denote by $\sigma(x)$ the
right cyclic shift of $x$, i.e. $\sigma(x)=(x_n,x_1,\ldots,x_{n-1})$.
Define recursively $\sigma^i(x)=\sigma(\sigma^{i-1}(x))$, for $i=2,3,\ldots$
and $\sigma^1(x)=\sigma(x)$. For $j<0$, we define
$\sigma^j(x)=\sigma^{\ell}(x)$, where $\ell = j \mod n$.

The next constructions are described in \cite{BRZ3}, although the dual codes
of the resulting family of $q$-ary completely regular codes are well known
as the family SU2 in \cite{Cald}.

Let $H$ be the parity check matrix of a $q$-ary cyclic Hamming code ${\mathcal H}_m$ of length $n=(q^m-1)/(q-1)$ (recall that a cyclic version of ${\mathcal H}_m$ exists exactly when $n$ and $q-1$ are coprime numbers). Clearly the simplex code generated by
$H$ is also a cyclic code.
For any $r\in \{1,2,\ldots,n\}$, consider the code $B^{(r)}$ of length $n r$
with the following parity check matrix $H_b(r)$ (we call it
Construction I in \cite{BRZ3}):
\begin{equation}\label{eq:3.1}
H_b(r) = \left[
\begin{array}{cccc}
H\;  & \;H \;  &\cdots\,&\,H\,\\
H_1\;& \;H_2 \;&\cdots\,&\,H_r
\end{array}
\right],\;\;r=1, \ldots, n;
\end{equation}
where $H_i$, for $1\leq i\leq n$, is the matrix $H \xi^i$ (here
$\xi\in\F_{q^m}$ is a primitive $n$th root of unity),
hence $H_i$ is obtained from $H$ by cyclically shifting $i$ times
its columns to the right (in these terms $H =H_n$). In \cite{BRZ3} we also presented Construction II. In this case, the
corresponding code of length $n(r+3)$, which we denote $A^{(r)}$, has
parity check matrix $H_a(r)$ of the form:
\begin{equation}\label{eq:3.2}
H_a(r) = \left[
\begin{array}{ccccccc}
H\;&\;O\:&\; H\;& \;H\;  & \;H \;  &\cdots\,&\,H\,\\
O\;&\;H\:&\; H\;&\; H_1\;& \;H_2 \;&\cdots\,&\,H_r
\end{array}
\right]\,.
\end{equation}
For the codes $A^{(r)}$ we fix the following interval for $r$:
\,$r \in \{-2, -1, 0, 1, \ldots, n-1\}$ (for $r=n$ we would have pairs of linear dependent columns), where $A^{(0)}$, $A^{(-1)}$
and $A^{(-2)}$ are defined by the parity check matrices, respectively
\begin{equation}\label{eq:3.20}
H_a(0) = \left[
\begin{array}{ccccccc}
H\;&\;O\:&\; H\\
O\;&\;H\:&\; H
\end{array}
\right]\,,\;\;
H_a(-1) = \left[
\begin{array}{ccccccc}
H\;&\;O\\
O\;&\;H
\end{array}
\right]\,,\;\;
H_a(-2) = \left[
\begin{array}{ccccccc}
H\\
O
\end{array}
\right]\,.
\end{equation}
Note that the codes $B^{(1)}$ and $A^{(-2)}$ are the Hamming code of
length $(q^m-1)/(q-1)$.

\begin{lemma}\label{equivBr}
Let $i$ and $j$ be positive integers such that $i+j\leq n$. The code $D$ with parity check matrix
\begin{equation}
\left[
\begin{array}{cccc}
H\;  & \;H \;  &\cdots\,&\,H\,\\
H_i\;&\;H_{i+1}\:&\cdots&\; H_{i+j-1}
\end{array}
\right]
\end{equation}
is equivalent to the code $B^{(j)}$.
\end{lemma}

\pr
The statement is evident, since the parity check matrices for the codes
$D$ and $B^{(j)}$ differ from each other by the multiplier $\xi^{i-1}$, applied to the second bottom part of rows.
\qed


In \cite{BRZ3} we proved the following theorem.

\begin{theorem}[\protect{\cite{BRZ3}}]\label{theo:3.1}
$\mbox{ }$
\begin{itemize}
\item [(i)] For any $r$,\,$2 \leq r \leq n$, the code $B^{(r)}$ with parity check matrix $H_b(r)$ given in (\ref{eq:3.1}) is a completely regular
$[nr,nr-2m,3;2]_q$-code with intersection array
$$
\IA=\{(q-1)nr,((q-1)n-r+2)(r-1);1,r(r-1)\}.
$$
\item[(ii)]  For any $r$,\,$0 \leq r \leq n-1$, the code $A^{(r)}$ with parity check matrix $H_a(r)$ given in (\ref{eq:3.2})  is a  completely
regular $[(r+3)n,(r+3)n-2m,3;2]_q$-code with intersection array
$$
\IA = \{(q-1)n(r+3), ((q-1)n-1-r)(r+2); 1, (r+2)(r+3)\}.
$$
\end{itemize}
\end{theorem}

In \cite{BRZ4} we defined the concept of {\em supplementary codes}. Let $H_m$ be the parity check matrix of a $q$-ary Hamming code of length $n=(q^m-1)/(q-1)$. Consider a non-empty subset of $n_A < n$ columns of $H_m$ as a parity check matrix of a code $A$. Call $B$ the code that has as parity check matrix the remaining $n_B=n-n_A$ columns of $H_m$. The code $A$ is the supplementary code of $B$, and vice versa. We summarize the results of \cite{BRZ4} in the following theorem.

\begin{theorem}[\cite{BRZ4}]\label{suppl}
Let $A$ and $B$ be supplementary codes as defined above.
\begin{itemize}
    \item[(i)] If $A$ has covering radius $1$, then it is a Hamming code. Both codes $A$ and $B$ are completely regular and completely transitive.
    \item[(ii)] If $A$ has covering radius $2$, dimension $n_A-m$, and $A$ is completely regular, then $B$ is completely regular with covering radius at most $2$.
    \item[(iii)] If $A$ has covering radius $2$, dimension $n_A-m$, and $A$ is completely transitive, then $B$ is completely transitive.
\end{itemize}
\end{theorem}

In all cases, the parameters and intersection arrays are computed in $\cite{BRZ4}$.

Now, consider the parity check matrix $H_{2m}$ of a $q$-ary Hamming code ${\mathcal H}_{2m}$ of length $(q^{2m}-1)/(q-1)$, such that the $q$-ary Hamming code ${\mathcal H}_{m}$ is cyclic. Equivalently, $n=(q^m-1)/(q-1)$ and $q-1$ are coprime numbers. We assume that $m\geq 3$ to avoid trivial cases. In fact, for $m=2$, ${\cal H}_m$ is cyclic only if $q$ is a power of $2$. Present $H_{2m}$ as follows:
\begin{equation}\label{matrixH2m}
H_{2m} = \left[\begin{array}{c|c}
H_b(r) & H_c(r)\\
\end{array}\right],\;\;\;r\in\{1,\ldots,q^m\},
\end{equation}
where $H_b(r)$ is the matrix (\ref{eq:3.1}) with $rn=r(q^m-1)/(q-1)$ columns, hence it is a parity check matrix for $B^{(r)}$. Call $C^{(r)}$ the supplementary code of $B^{(r)}$, i.e. $C^{(r)}$ has parity check matrix $H_c(r)$. By combining the results of \cite{BRZ3} and \cite{BRZ4}, we obtain the following theorem.

\begin{theorem}\label{params}
Let $n=(q^m-1)/(q-1)$ and let $B^{(r)}$ and $C^{(r)}$ be the supplementary codes as defined above.
\begin{itemize}
    \item[(i)] $B^{(1)}$ is a Hamming $[n,n-m,3;1]_q$-code and $C^{(1)}$ is a $[q^mn, q^mn-2m,3,2]_q$-code. Both codes are completely regular and completely transitive with intersection arrays
    \begin{eqnarray*}
    \IA(B^{(1)}) &=& \{q^m-1;1\};\\
    \IA(C^{(1)}) &=& \{q^m(q^m-1), q^m-1; 1, q^m(q^m-1)\}.
    \end{eqnarray*}
    \item[(ii)] For $2\leq r\leq n$, $B^{(r)}$ is a $[rn,rn-2m,3;2]_q$-code and $C^{(r)}$ is a\\ $[(q^m+1-r)n,(q^m+1-r)n-2m,3;2]_q$-code. Both codes are completely regular with intersection arrays
    \begin{eqnarray*}
    \IA(B^{(r)}) &=& \{r(q^m-1),(r-1)(q^m+1-r);1,r(r-1)\};\\
    \IA(C^{(r)}) &=& \{(q^m+1-r)(q^m-1),r(q^m-r); 1,(q^m+1-r)(q^m-r) \}.
    \end{eqnarray*}
    Furthermore, $B^{(r)}$ is completely transitive if and only if $C^{(r)}$ is completely transitive.
\end{itemize}
\end{theorem}

\begin{remark}
In the binary case, $q=2$, the matrix $H_{2m}$ can be written as
\begin{equation}
H_{2m}=\left[
\begin{array}{cccc|cccccc}
H\;  & \;H \;  &\cdots\,&\,H\,& H &\cdots & H & H & \mathbf{0} & H\\
H_1\;&\;H_2\:&\cdots&\; H_r   & H_{r+1} & \cdots & H_{n-1} & \mathbf{0} & H & H
\end{array}
\right],
\end{equation}
where $H$ is a parity check matrix of a binary cyclic Hamming code $\mathcal{H}_m$.

Indeed, $H_{2m}$ has $n(n+2)=2^{2m}-1$ columns and there are no repeated columns. Therefore, in the binary case and by Lemma \ref{equivBr}, the code $C^{(r)}$ is equivalent to the code $A^{(n-r-1)}$ (Construction II in \cite{BRZ3}).
\end{remark}

\section{Completely transitive families}\label{sec:3}

For the rest of the paper, let $B^{(r)}$ be the code with parity check matrix $H_b(r)$ given in (\ref{eq:3.1}). Let $T_1,\ldots,T_r$ be the $n$-sets, which we call {\em blocks}, of coordinate positions corresponding to the columns of $H_1,\ldots,H_r$. That is,
$$
T_j=\{(j-1)n+1,(j-1)n+2,\ldots,jn\},\;\;\;\forall j=1,\ldots,r.
$$
Consider also the vectors of weight one indexed as follows:
$$
e_1^1,\ldots,e_n^1, e_1^2,\ldots,e_n^2,\ldots,\ldots,e_1^r,\ldots,e_n^r;
$$
where the superindices indicate the corresponding block.

In order to establish in which cases the codes of Theorem \ref{params} are completely transitive, we need several previous results.

\begin{lemma}\label{not2Hom}
The general linear group $\GL(m,q)$ ($m\geq 3$) is not $2$-homogeneous, unless $q=2$.
\end{lemma}

\pr
It is well known that the order of $\GL(m,q)$ is
$\prod_{k=0}^{m-1} (q^m-q^k)$,
which is even. Hence, if $\GL(m,q)$ is $2$-homogeneous, it is also $2$-transitive by \cite[Lemma 2.1]{Sha}, implying $q=2$.
\qed

\begin{lemma}\label{cosets2}
Let $n=(q^m-1)/(q-1)$. For $1< r \leq n$, let $B^{(r)}$ be the $[nr,nr-2m,3;2]_q$-code obtained by Construction I. Then, the number of cosets of $B^{(r)}$ with minimum weight 2 is $q^{2m}-1-r(q^m-1)$.
\end{lemma}

\pr
Since $B^{(r)}$ has minimum distance $3$, we have that the number of cosets of weight $1$ is $rn(q-1)=r(q^m-1)$. The total number of cosets is $q^{rn}/(q^{rn-2m})=q^{2m}$. Hence, the number of cosets with minimum weight greater than $1$ is $q^{2m}-1-r(q^m-1)$. All such cosets have minimum weight $2$, because the covering radius of $B^{(r)}$ is $\rho=2$.
\qed

\begin{lemma}\label{lem:3.3}
Let $B^{(r)}$ be the $[nr,nr-2m,3;2]_q$-code obtained by Construction I, where $n=(q^m-1)/(q-1)$. Then,
\begin{itemize}
\item[(i)] $\MAut(B^{(1)}) = \GL(m,q)$.
\item[(ii)] $\MAut(B^{(2)}) = \GL(m,q) \times \GL(m,q)\times C_2$.
\item[(iii)] $\MAut(B^{(3)}) = \GL(m,q) \times S_3$.
\end{itemize}
\end{lemma}

\pr
The statement (i) is trivial, since $B^{(1)}$ is a Hamming code of length $(q^m-1)/(q-1)$. It is well known that its automorphism group is $\GL(m,q)$ \cite{MacW}.

For (ii), note that, after linear operations on the rows, the parity check matrix $H_b(2)$ of the code $B^{(2)}$ can be presented in the form
\[
H_b(2) = \left[
\begin{array}{c|c}
H \,&\, O \\
O \,& \, H
\end{array}
\right].
\]
Since the Hamming code of length $(q^m-1)/(q-1)$ (defined by the parity check matrix $H$) is stabilized by the group $\GL(m,q)$, we have that the code $B^{(2)}$ is invariant under the action of any matrix $G$ of type
\begin{equation}\label{matrixii}
G = \left[
\begin{array}{c|c}
G_1 \,&\, O \\
O \,& \, G_2
\end{array}
\right],
\end{equation}
where both matrices $G_1$ and $G_2$ are arbitrary matrices from the group $\GL(m,q)$. We conclude that the monomial automorphism group $\MAut(B^{(2)})$ contains as a subgroup the group $\GL(m,q) \times \GL(m,q)$.
From the other side, we can clearly interchange the blocks corresponding to the columns of $G_1$ and $G_2$, respectively, without changing the code $B^{(2)}$. Hence the group $\MAut(B^{(2)})$ contains as a subgroup the group of order $2$. It is easy to see that any monomial automorphism of $B^{(2)}$ contains only automorphisms from these groups above. Indeed, assume that the matrix $G$ belongs to $\GL(2m,q)$, but not to $\GL(m,q) \times \GL(m,q)$. Then we can see (by solving the corresponding system of linear equations) that any column of $H_b(2)$ can be moved to any other column. But this leads to a contradiction to the shape of $H_b(2)$. Hence, we conclude that $\MAut(B^{(2)})$ contains as subgroups only $\GL(m,q) \times \GL(m,q)$
and $C_2$. Since the two cosets of
$\GL(m,q) \times \GL(m,q)$ induced by $C_2$ coincide, we deduce that $\MAut(B^{(2)})$ is a direct product of all three subgroups, i.e.
\[
\MAut(B^{(2)}) = \GL(m,q) \times \GL(m,q) \times C_2\,.
\]

For (iii), the parity check matrix $H_b(3)$ (see (\ref{eq:3.1})) of $B^{(3)}$ can be transformed, by linear operations on the rows and column permutations, to the following form:
\[
H_b(3) \simeq \left[
\begin{array}{c|c|c}
H \,&\, O \,&\,H\\
O \,& \,H\, &\,H
\end{array}
\right].
\]
The exclusive property of this matrix is that any of all three blocks is stabilized by all matrices $G$ of the type
\[
G = \left[
\begin{array}{c|c}
G_1 \,&\, O \\
O \,& \, G_1
\end{array}
\right],
\]
where $G_1$ is an arbitrary matrix from the group $\GL(m,q)$. We conclude that the group $\MAut(B^{(3)})$ contains as a subgroup the group $\GL(m,q)$. Now we can see that any codeword $z = (x, y, u)$ of $B^{(3)}$ can be presented in the form $z = (u+x_1, u+x_2, u+x_3)$, where $x_i$ is a codeword of the Hamming code (of length $n=(q^m-1)/(q-1)$, i.e. it has the support only in the block $T_i$, \,$i=1,2,3$, and
$u$ is an arbitrary binary vector of length $n$). We deduce that all three blocks can be arbitrary permuted without changing of the code $B^{(3)}$ and, therefore, $\MAut(B^{(3)})$ contains as a subgroup the group $S_3$. Similarly to the previous case, we can see that these are the only monomial automorphisms of the code $B^{(3)}$.
Indeed, if for example $G_1 \neq G_2$, where $G$ is of the form (\ref{matrixii}) (see case (ii)), we change the third block of $H_b(3)$ and obtain the column which does not belong to the third block of $H_b(3)$. Since any element of $\GL(m,q)\times \GL(m,q)$ cannot interchange the blocks and since they intersect only in the identity element,
we conclude that the group $\MAut(B^{(3)})$
is the direct product of groups $\GL(m,q)$ and $S_3$, i.e.
\[
\MAut(B^{(3)}) =\GL(m,q) \times S_3\,.
\]

Evidently, the actions of both $\GL(m,q)$ and $S_3$ do not depend on each other.
\qed

The next statement shows cases in which the code $B^{(r)}$ is
completely transitive (hence, so is $C^{(r)}$).

\begin{proposition}\label{CTcodes}
The codes $B^{(1)}$ and $B^{(2)}$ are completely transitive.
Furthermore, for $q=2$, the codes $B^{(3)}$, $B^{(n-1)}$ and
$B^{(n)}$ are also completely transitive (here $n=(q^m-1)/(q-1)$).
\end{proposition}

\pr
The code $B^{(1)}$ is a Hamming code. Its automorphism group is $\GL(m,q)$ \cite{MacW}, that acts transitively over the set of vectors of weight one. Hence all cosets with minimum weight one are in the same orbit and thus $B^{(1)}$ is completely transitive.

For $B^{(2)}$, we have that its automorphism group is $\GL(m,q)\times \GL(m,q)\times C_2$ (Lemma \ref{lem:3.3}) which acts transitively over the set of vectors of weight one in each block and also both blocks can be interchanged. Therefore all cosets with minimum weight one are in the same orbit. By Theorem \ref{theo:3.1}, the covering radius of $B^{(2)}$ is $2$. Hence, we only need to prove that the cosets with minimum weight two are also in the same orbit. Note that any vector at distance $2$ from the code has its nonzero coordinates in different blocks of coordinate positions. Since we can act to these blocks independently, we have that the cosets with minimum weight two of $B^{(2)}$ are in the same orbit.

For $q=2$, by Lemma \ref{lem:3.3}, the automorphism group of $B^{(3)}$ is $\GL(m,2) \times S_3$, that clearly acts
transitively over the coordinate positions. Hence, the cosets of $B^{(3)}$ with minimum weight one are in the same orbit. Consider two cosets of $B^{(3)}$ with minimum weight two, $B^{(3)}+\bx$ and $B^{(3)}+\by$, where $\bx=e_i^a+e_j^b$ and $\by=e_k^c+e_\ell^d$. Note that
$$
i\neq j, k\neq \ell, a\neq b, c\neq d;
$$
otherwise both vectors $\bx$ and $\by$ would not be at distance $2$ from the code. Since $\GL(m,2)$ is doubly transitive \cite{MacW}, there exists $\phi\in\MAut(B^{(3)})$ such that $\phi(e_i^a+e_j^b)=e_k^a+e_\ell^b$. Now, consider $\theta\in S_3$ such that $\theta(T_a)=T_c$ and $\theta(T_b)=T_d$. We obtain that $\theta(\phi(\bx))=\by$. Therefore all cosets with minimum weight $2$ are in the same orbit. Thus, $B^{(3)}$ is completely transitive since its covering radius is $2$.

For $q=2$, the supplementary codes of $B^{(n-1)}$ and $B^{(n)}$ have, respectively, parity check matrices $H_a(0)$ and $H_a(-1)$ defined in (\ref{eq:3.20}). We have seen that these codes are equivalent to $B^{(3)}$ and $B^{(2)}$, respectively. From Theorem \ref{params}, we deduce that $B^{(n-1)}$ and $B^{(n)}$ are completely transitive codes.
\qed

\begin{lemma}\label{B3noCT}
The code $B^{(3)}$ is completely transitive if and only if $q=2$.
\end{lemma}

\pr
By Proposition \ref{CTcodes} we know that $B^{(3)}$ is completely
transitive for $q=2$. We have to prove that it is not for $q>2$.

The code $B^{(3)}$ is equivalent to the code $C$ that has parity check matrix:
$$
H_b(3)=\left[
\begin{array}{c|c|c}
H \,&\, O \,&\,H\\
O \,& \,H\, &\,H
\end{array}
\right].
$$
Consider the set of vectors
$$
S=\{\alpha e_i^1 + e_j^2\mid (i,\alpha)\neq (j,1),\;\; \alpha\in\F_q\setminus\{0\},\;\; i,j\in\{1,\ldots,n\}\}.
$$
Clearly, any vector in $S$ is at distance two from $C$. Indeed, if $\bx=\alpha e_i^1 + e_j^2\in S$, then $H_b(3)\bx^T$ is a nonzero vector which is not a column of $H_b(3)$. Moreover, given two such vectors $\bx=\alpha e_i^1 + e_j^2$ and $\by=\alpha' e_{i'}^1 + e_{j'}^2$, where $i\neq i'$ or $j\neq j'$ or $\alpha\neq\alpha'$, we have that $H_b(3)(\bx-\by)^T$ is a nonzero vector, thus $\bx-\by\notin C$. Therefore, all vectors in $S$ are in different cosets of $C$.

Compute the cardinality of $S$:
\begin{itemize}
    \item[(i)] Vectors of the form $\alpha e_i^1+ e_i^2$, ($\alpha\neq 1$). There are $n(q-2)$ such vectors.
    \item[(ii)] Vectors of the form $\alpha e_i^1 +  e_j^2$, where $i\neq j$. There are $n(q-1)(n-1)$ such vectors.
\end{itemize}
Hence,
$$
|S|=n(q-2)+n(q-1)(n-1)=n(q^m-2)=\frac{(q^m-1)(q^m-2)}{q-1}.
$$

As a consequence, we can find $(q^m-1)(q^m-2)/(q-1)$ vectors
which are all in different cosets of minimum weight $2$. Thus,
if all such cosets are in the same orbit, we have that the
monomial automorphisms in $\GL(q,m)$ of the form
\[ g=
\left[
\begin{array}{cc}
g_1 & O \\
O   & g_1
\end{array}
\right]
\]
acts $2$-homogeneously on the first two blocks (indeed, for any
vector $\alpha e_i^1+ \beta e_j^2$, we can consider a multiple
$\alpha' e_i^1+ e_j^2$ in the same coset). But this implies
that the action of $\GL(q,m)$ in only one block is $2$-homogeneous,
since any automorphism like $g$ acts identically in both blocks.
Thus, we have $q=2$, by Lemma \ref{not2Hom}.
\qed

\begin{remark}
According to Lemma \ref{cosets2}, the total number of cosets of
minimum weight $2$ is $(q^m-1)(q^m-2)$. Note that for $q=2$, the
number of cosets induced by the set $S$ coincides and since the
action of $\GL(q,m)$ is doubly transitive, all the cosets of
minimum weight $2$ are in the same orbit. This is an alternative
argument to see that $B^{(3)}$ is completely transitive in the
binary case (Proposition \ref{CTcodes}).
\end{remark}

To determine the monomial automorphism group of $B^{(r)}$ in the general case seems to be a challenging problem. However, many computational results using {\sc magma} \cite{Magma}, suggest that the order of the monomial automorphism group is not greater than $8n(q-1)$, for $r>3$ except when $q=2$ and $r\in\{n-1,n\}$. In these last two cases we have $\MAut(B^{(n-1)}) =\MAut(B^{(3)}) = \GL(m,2) \times S_3$ and  $\MAut(B^{(n)}) =\MAut(B^{(2)}) = \GL(m,2) \times \GL(m,2) \times C_2$, respectively, and it is clear that in these two cases the order of $\MAut(B^{(r)})$ is greater than $8n(q-1)$.
Assuming such upper bound we have that the codes not mentioned in Proposition \ref{CTcodes} are not completely transitive.

\begin{proposition}\label{fita}
If $B^{(r)}$ is completely transitive, then $|\MAut(B^{(r)})|=cn(q-1)$, where
$$
c\geq \max\{r,n(q-1)+2-r\}.
$$
\end{proposition}

\pr
  If $B^{(r)}$ is completely transitive, then $\MAut(B^{(r)})$ must be transitive over the set of vectors of weight one since all cosets of minimum weight one must be in the same orbit. Hence $|\MAut(B^{(r)})|$ is a multiple of $rn(q-1)$, say $cn(q-1)$. Moreover, $cn(q-1) \geq rn(q-1)$ since $B^{(r)}$ has $rn(q-1)$ cosets of minimum weight one. By Lemma \ref{cosets2}, the number of cosets of minimum weight two is $n(q-1)(n(q-1)+2-r)$. All such cosets must be in the same orbit and thus $c \geq n(q-1)+2-r$.
\qed

\begin{corollary}\label{NoCT}
Assume that $|\MAut(B^{(r)})| \leq 8n(q-1)$, for $r>3$ except when $q=2$ and $r\in\{n-1,n\}$.
\begin{itemize}
\item[(i)] If $q=2$ and $4\leq r < n-1$, then $B^{(r)}$ is not completely transitive.
\item[(ii)] If $q>2$ and $4\leq r \leq n$, then $B^{(r)}$ is not completely transitive.
\end{itemize}
\end{corollary}

\pr
From Proposition \ref{fita}, we have $r\leq 8$ and $n(q-1)+2-r \leq 8$. Combining both inequalities, we obtain $n(q-1)\leq 14$. Thus,

(i) If $q=2$, then $m\leq 3$ (and $n \leq 7$). Hence, the only possible cases are $m=3$, $r\in\{4,5\}$.

(ii) If $q>2$, then the only possible cases are $q=3$, $m=2$ (hence $n=4$), $r\in\{3,4\}$.

For these remaining cases, $(q,m,r)\in\{(2,3,4),(2,3,5),(3,2,3),(3,2,4)\}$, we have verified with {\sc magma} that the corresponding codes are not completely transitive.
\qed

\bigskip

From Proposition \ref{CTcodes}, Lemma \ref{B3noCT}, and Corollary \ref{NoCT} we can state the main result of this paper.

\begin{theorem}\label{ctc}
Let $B^{(r)}$ be the code over $\F_q$ of length $rn=r(q^m-1)/(q-1)$ ($1\leq r\leq n$), obtained by Construction I. Let $C^{(r)}$ be the supplementary code of $B^{(r)}$, such that the concatenation of the parity check matrices of $B^{(r)}$ and $C^{(r)}$ is $H_{2m}$, the parity check matrix of a Hamming code of length $(q^{2m}-1)/(q-1)$. Then,
\begin{itemize}
\item[(i)] For $q = 2$, the codes $B^{(r)}$ and $C^{(r)}$ are completely transitive if $r \in \{1, 2, 3, n-1, n\}$.
\item[(ii)] For $q > 2$, the codes $B^{(r)}$ and $C^{(r)}$ are completely transitive if $r \in \{1, 2\}$.
\item[(iii)] For $q>2$, the codes $B^{(3)}$ and $C^{(3)}$ are not completely transitive.
\item[(iv)] For the remaining cases (not in (i), (ii) nor (iii)) the codes $B^{(r)}$ and $C^{(r)}$ are not completely transitive, assuming that in these cases the order of $\MAut(B^{(r)})$ is not greater that $8n(q-1)$.
\end{itemize}
\end{theorem}

\section*{Acknowledgements}

This work has been partially supported by the Spanish Grant
PID2022-137924NB-I00 (AEI/10.13039/501100011033) and the Catalan
AGAUR Grant 2021SGR-00643. The research of the third author of the
paper was carried out at the IITP RAS within the program of
fundamental research on the topic ``Mathematical Foundations of
the Theory of Error-Correcting Codes" and was also supported by
the National Science Foundation of Bulgaria under the project no.
20-51-18002.

\end{document}